\documentclass[aps,prstab,twocolumn,showpacs]{revtex4}
\usepackage{graphics}
\usepackage{bm}

\begin{document}

\title{COUPLING IMPEDANCES OF SMALL DISCONTINUITIES: \\
 DEPENDENCE ON BEAM VELOCITY} 
\author{Sergey~S.~Kurennoy}
\affiliation{Los Alamos National Laboratory,
         Los Alamos, NM 87545, USA}
\pacs{41.75.-i,41.20.-q}

\begin{abstract} 
The beam coupling impedances of small discontinuities of an 
accelerator vacuum chamber have been calculated 
[e.g., S.S.~Kurennoy, R.L.~Gluckstern, and G.V.~Stupakov, 
Phys.\ Rev.\ E {\bf 52}, 4354 (1995)] 
for ultrarelativistic beams using the Bethe diffraction 
theory. Here we extend the results to an arbitrary beam velocity. 
The vacuum chamber is assumed to have an arbitrary, but uniform 
along the beam path, cross section. The longitudinal and transverse 
coupling impedances are derived in terms of series over 
cross-section eigenfunctions, while the discontinuity shape 
enters via its polarizabilities. Simple explicit formulas for two 
important particular cases --- circular and rectangular chamber 
cross sections --- are presented. The impedance dependence 
on the beam velocity exhibits some unusual features: for example, 
the reactive impedance, which dominates in the ultrarelativistic 
limit, can vanish at a certain beam velocity, or its magnitude can 
exceed the ultrarelativistic value many times. In addition, we 
demonstrate that the same technique, the field expansion into a 
series of cross-section eigenfunctions, is convenient for 
calculating the space-charge impedance of uniform beam pipes 
with arbitrary cross section.
\end{abstract} 

\maketitle

\section{Introduction} 

A common tendency in design of modern accelerators is to minimize 
beam-chamber coupling impedances to avoid beam instabilities and 
reduce heating. Even contributions from tiny discontinuities
like pumping holes have to be accounted for because of their 
typically large numbers. Direct numerical methods encounter 
difficulties in calculating the impedances of small discontinuities.
The difficulties of usual time-domain methods --- computing
wake potentials and then finding the impedances as their
Fourier transforms --- are mostly technical for ultra\-relativistic 
beams: one has to apply very fine steps in wake computations to 
resolve small obstacles. However, for non-ultra\-relativistic
beams the usual numerical approach fails due to difficulty
in implementing proper boundary conditions at the open ends 
of the beam pipe. This situation makes analytical methods 
especially important. A general analytical approach for calculating 
the beam coupling impedances of small discontinuities 
on the walls of an accelerator vacuum chamber has been developed 
in \cite{SK92,KGS} for ultra\-relativistic beams. 
The method is based on the Bethe theory of diffraction by small 
holes \cite{Bethe}, according to which the fields diffracted by a 
hole can be found as those radiated by effective electric and 
magnetic dipoles. 

The Bethe idea of effective dipoles developed in 1944 was first 
used to calculate the coupling impedances of pumping holes in a 
circular waveguide for the case of an ultra\-relativistic beam 
in 1992 \cite{SK,RLG}. 
Bethe's theory is applicable for wavelengths large compared to 
the typical hole size $h$, and it can be used for impedance 
calculations when $h$ is much smaller than the typical dimension 
$b$ of the chamber cross section. Within this limitation, the 
theory was applied to vacuum chambers with an arbitrary 
simply-connected cross section in \cite{SK92}, again for 
relativistic beams. The imaginary part of the impedance was shown 
to be proportional to the difference of hole polarizabilities 
$(\psi -\chi)$, where the magnetic susceptibility $\psi$ and the 
electric polarizability $\chi$ are both small compared to $b^3$. 
From considerations of the energy radiated into the chamber and 
through the hole, the real part of the hole impedance comes out to 
be proportional to $(\psi^2 +\chi^2)$, being usually much smaller 
than the reactance. These results are not restricted to small holes, 
but remain valid for other small discontinuities like posts, masks, 
or irises, because the idea of effective polarizabilities works 
equally well in those cases, as was demonstrated in \cite{K&S,SK97}.
A more refined theory that takes into account the reaction 
of radiated waves back on the hole was developed in \cite{KGS}.
In the latter approach, the beam coupling impedances of a 
discontinuity come out as perturbative series in polarizabilities, 
more exactly, in small parameters $\psi/b^3$ and $\chi/b^3$: 
the reactive impedance is the first order effect, while the 
real part of the impedance has the second order. Moreover, 
the theory \cite{KGS} contains non-perturbative effects: it 
gives the trapped modes near the pipe cut-off frequencies 
due to some small discontinuities. This effect was discovered 
earlier using a different method \cite{S&K}.

In the present article we extend the analytical approach 
\cite{SK92,KGS} to the case of non-ultra\-relativistic beams.  
The beam coupling impedances of a small discontinuity on the walls
of a vacuum chamber with any simply-connected cross section are 
derived for an arbitrary beam velocity $v=\beta c$. 
The theory gives analytical expressions for 
the longitudinal and transverse coupling impedances 
in terms of series over cross-section eigenfunctions. 
The shape of the discontinuity enters via its electric and 
magnetic polarizabilities. Previous results for 
ultrarelativistic beams are naturally reproduced in the 
limit of $\beta \to 1$. Simple explicit expressions for the 
impedances of non-ultrarelativistic beams are derived for 
circular and rectangular chamber cross sections.  

One should mention a few earlier results related to the subject. 
The coupling impedances of a small hole in a circular beam pipe
were calculated \cite{Palumbo96} for $\beta < 1$ in the form of 
series involving roots of Bessel functions. 
Closed-form expressions of the hole impedances for 
non-relativistic beams in a cylindrical vacuum chamber with a 
circular cross section were obtained by Gluckstern {\it et al}
\cite{G&F99,RLG2000}. The longitudinal impedance of a 
small round hole in a circular waveguide was derived 
in a closed form for a beam with $\beta < 1$ and a finite 
transverse size in \cite{Al-Kh01}. In a somewhat related
paper \cite{SK99losbet}, the bunch loss factors were 
investigated for non-ultra\-relativistic beams.
All the results above, however, have been restricted to 
the particular case of an axisymmetric vacuum chamber. 
Our study treats a more general case of an arbitrary 
chamber cross section.

In addition, we show that the same technique --- based on 
field expansions in cross-section eigenfunctions --- works 
very efficiently for deriving the space-charge impedance 
of a uniform waveguide having an arbitrary  
cross section. A general expression is obtained that includes
frequency corrections to and beam-velocity dependence of the 
space-charge $g$-factor. In the particular case of a circular 
cross section, the results coincide with the known ones, e.g.\ 
\cite{RLG2000}. For a rectangular chamber, the space-charge 
impedance is expressed in the form of a simple series 
convenient for computation.   

The paper is organized as follows. A general analysis of the beam 
fields and fields scattered by a small discontinuity in the vacuum 
chamber is given in Sec.~II. 
Section~III presents the derivation and results for the 
coupling impedances of small discontinuities. 
In Sec.~IV the space-charge impedance of a homogeneous chamber 
with an arbitrary cross section is calculated. 
The eigenfunctions and some derivations for the two particular 
cases of the vacuum chamber --- with a circular and rectangular 
cross section --- are presented in the Appendices. 

\section{Fields}

Let us consider an infinite cylindrical pipe with an arbitrary 
cross section $S$ and perfectly conducting walls. The $z$ axis is
directed along the pipe axis, a small discontinuity (e.g., a hole) 
is located in the cross section $z=0$ at the point ($\vec{b},0$), 
and a typical hole size $h$ satisfies $h\ll b$. 
The discontinuity is considered small 
when its size is much smaller than the wavelength of interest.  
To evaluate the coupling impedance one has to calculate the 
fields induced in the chamber by a given current. Consider a 
charge $q$ that moves on or parallel to the chamber axis  
with velocity $v=\beta c$. A rigid bunch is usually assumed for 
wake or impedance calculations. It is convenient to choose a 
"pancake" charge distribution such that the charge and current 
densities are given by 
\begin{eqnarray}
 \rho(\vec{r},z;t) & = & q f(\vec{r}\,) 
 \delta(z-\beta c t) \ ,                 \nonumber \\ 
 \vec{j}(\vec{r},z;t) & = & q \beta c f(\vec{r}\,) 
  \delta(z-\beta c t) \hat{z}            \ , \label{chrgt}      
\end{eqnarray}
where $f(\vec{r}\,)$ is a normalized transverse charge distribution,
$\int_S d\vec{r} f(\vec{r}\,) = 1$, $\delta(x)$ is Dirac's delta 
function, and $\hat{z}$ is a unit vector in the $z$-direction. 
The distribution (\ref{chrgt}) includes two particular cases that
are especially convenient in calculations: a point charge with 
the transverse offset $\vec{s}$ from the axis, when
$f(\vec{r}\,)=\delta(\vec{r}-\vec{s})$, 
and a thin axisymmetric ring of radius $a$ with 
$f(\vec{r}\,)=\delta(r-a)/(2\pi a)$.
In frequency domain, omitting factor $\exp(-i\omega t)$, 
the current density harmonic (\ref{chrgt}) becomes
\begin{equation} 
 \vec{j}(\vec{r},z;\omega) = q f(\vec{r}\,) 
 e^{i\omega z/(\beta c)} \hat{z} \ .  \label{currw}      
\end{equation}
This harmonic corresponds to the space harmonic 
$k=\omega/(\beta c)$. From Maxwells's equations in frequency 
domain follow the wave equations for harmonics of the beam 
field components:
\begin{eqnarray}
 \left(\nabla^2+\partial^2_z \right) E_z + 
 \omega^2/c^2 E_z & = & i \omega \mu_0 /
 (\beta^2 \gamma^2) j_z                     \,, \nonumber \\
 \left(\nabla^2+\partial^2_z \right) \vec{E}_\bot + 
 \omega^2/c^2 \vec{E}_\bot & = & 1/
 (\varepsilon_0 \beta c) \vec{\nabla}j_z \,, \label{fldw} \\ 
\left(\nabla^2+\partial^2_z \right) \vec{H}_\bot + 
 \omega^2/c^2 \vec{H}_\bot & = & 
 -\left(\hat{z}\times\vec{\nabla}\right)j_z \,. \nonumber       
\end{eqnarray}
Here all quantities are functions of $(\vec{r},z;\omega)$,
$\vec{\nabla}$ is the two-dimensional (2D) gradient in plane 
$S$, $\partial_z \equiv \partial/\partial z$, 
$\gamma=1/\sqrt{1-\beta^2}$, and $j_z$ is defined 
from Eq.~(\ref{currw}). The longitudinal component of 
the magnetic field is not excited
by the current (\ref{chrgt}), so that $H_z=0$. 

Let us now introduce eigenvalues $k^2_g$ and ortho\-norma\-lized 
eigenfunctions (EFs) $e_g(\vec{r})$ of the Dirichlet 
boundary problem in $S$:
\begin{equation}
\left (\nabla ^2+ k^2_g\right ) e_g = 0 \ ; 
  \qquad e_g\big\vert_{\partial S} = 0 \ , \label{boundpr}
\end{equation}
where $g=\{n,m\}$ is a generalized 2D index. Since 
EFs form a complete set in $S$, the transverse dependence of 
solutions to Eqs.~(\ref{fldw}) can be found as a series in EFs.
Obviously, the solutions should depend on the longitudinal 
coordinate as $\exp(i \omega z/\beta c)$.
Then the fields harmonics $\vec{E},\vec{H}$ produced by 
the charge distribution (\ref{chrgt}) at the location 
$(\vec{b},z)$ on the chamber wall without hole can be 
expressed in terms of EFs (\ref{boundpr}) as
\begin{eqnarray}
 E_{\nu} (\vec{b},z;\omega) & = & 
 Z_0H_{\tau}(\vec{b},z;\omega)/\beta           \label{beamf}   
 \\ & = & - \frac{Z_0q}{\beta} 
 \exp{\left(i\frac{\omega z}{\beta c}\right)}
 \sum_g \frac{f_g \nabla _{\nu}e_g(\vec{b}\,)} 
  {k^2_g + \kappa^2}  \ ,  \nonumber       
\end{eqnarray}
where it was convenient to introduce  
$\kappa \equiv \omega/(\beta\gamma c)$.
Here $Z_0 =\sqrt{\mu_0/\varepsilon_0} = 120 \pi$~Ohms is the 
impedance of free space, $\hat{\nu}$ means an outward normal unit 
vector, $\hat{\tau}$ is a unit vector tangent to the boundary 
$\partial S$ of the chamber cross section $S$,
$\nabla _{\nu} \equiv \vec{\nabla}\cdot\hat{\nu}$, and 
$\{ \hat{\nu},\hat{\tau},\hat{z}\}$ form a right-handed basis.
In Eq.~(\ref{beamf}) $f_g$ are the coefficients of EF expansion 
$f(\vec{r}\,) = \sum_g f_g e_g(\vec{r}\,)$; they are given by
$f_g=\int_S d\vec{r}\, f(\vec{r}\,) e_g(\vec{r}\,)$. For the case 
of a point charge with the transverse offset $\vec{s}$ from the 
axis, we have $f_g=e_g(\vec{s}\,)$. The eigenvalues and EFs for 
particular cross sections are given in the Appendix.

In a similar way, the longitudinal component of the electric 
field produced by the current (\ref{chrgt}) as a function of
the transverse coordinates $\vec{r}$ is 
\begin{equation}
 E_z(\vec{r},z;\omega) = - i \frac{\omega}{\beta c} 
 \frac{Z_0q}{\beta \gamma^2} 
 e^{i\omega z/(\beta c)} \sum_g \frac{f_g e_g(\vec{r}\,)} 
  {k^2_g + \kappa^2}  \ .                          \label{Ezf}      
\end{equation}
Obviously, $E_z$ vanishes at the chamber wall because of the 
Dirichlet boundary conditions (BCs) for EFs (\ref{boundpr}). 
The transverse fields on the wall (\ref{beamf}) are not zero 
since the series (\ref{beamf}) includes the gradients 
of EFs and $\nabla _{\nu}e_g(\vec{b}\,) \neq 0$.  

Note that one could use EFs different from EFs (\ref{boundpr}) 
to calculate the transverse magnetic field $\vec{H}_\bot$. 
A convenient choice would be to use EFs $h_g(\vec{r}\,)$ 
of the Neumann boundary problem, which differs from 
(\ref{boundpr}) by the BCs, 
$\nabla _{\nu}h_g\big\vert_{\partial S} = 0$. 
However, we chose to relate $\vec{H}_\bot$ to the transverse 
electric field $\vec{E}_\bot$ directly from the Maxwell equations. 
For the harmonic $k=\omega/(\beta c)$, one gets
$Z_0 \hat{z} \times \vec{H}_\bot = - \beta \vec{E}_\bot$,
from which the first line in Eqs.~(\ref{beamf}) follows 
immediately.

One more remark about Eqs.~(\ref{beamf}) is worthwhile. Since 
the lowest eigenvalue $k_g$ is of the order of $1/b$, in the 
long-wavelength limit, $\kappa b = \omega b/(\beta\gamma c) \ll 1$, 
the sums in Eqs.~(\ref{beamf})-(\ref{Ezf}) become independent of 
frequency and of $\beta$. The only remaining dependence on the 
beam velocity $E_\nu \propto 1/\beta$ seems counter-intuitive: one
would expect $H_\tau \propto \beta$. The apparent contradiction 
disappears as soon as we recall that Eqs.~(\ref{beamf}) give 
the field harmonics with wavenumber $k=\omega/(\beta c)$, 
not the beam fields in time domain.  

Now we have to calculate the fields scattered into the vacuum 
chamber by the discontinuity. According to the Bethe theory, 
at distances $l$ such that $h \ll l \ll b$, the fields radiated 
by the discontinuity (hole) into the pipe are equal to those 
produced by effective dipoles \cite{Bethe,Collin} 
\begin{eqnarray}
P_\nu & = & - \chi \varepsilon_0 E^h_\nu/2; \quad  
M_\tau = (\psi_{\tau \tau} H^h_\tau + \psi_{\tau z} H^h_z )/ 2; 
 \nonumber \\
M_z & = & (\psi_{z \tau} H^h_\tau + \psi_{z z} H^h_z )/ 2 
 \ ,                 \label{dip}
\end{eqnarray}
where superscript '$h$' means that the beam fields (\ref{beamf}) 
are taken at the hole location $(\vec{b},0)$.
Polarizabilities $\psi, \chi$ are related to the effective 
ones $\alpha_e, \alpha_m$ used in \cite{Collin,SK} as 
$\alpha_e=-\chi/2$ and $\alpha_m=\psi/2$, so that for a circular 
hole of radius $h$ in a thin wall $\psi=8h^3/3$ and $\chi=4h^3/3$ 
\cite{Bethe}. In general, $\psi$ is a symmetric 2D-tensor, 
which can be diagonalized. If the discontinuity is symmetric, and 
its symmetry axis is parallel to $\hat{z}$, the skew terms vanish, 
i.e.\ $\psi_{\tau z}=\psi_{z\tau}=0$. In a more general case of 
a non-zero tilt angle $\alpha $ between the major symmetry axis 
and $\hat{z}$, 
\begin{eqnarray}
 \psi_{\tau \tau} & = & \psi_\bot \cos^2{\alpha} + 
 \psi_{\|} \sin^2{\alpha} \, , \nonumber \\
 \psi_{\tau z} & = & \psi_{z \tau } = (\psi_{\|} 
 - \psi_\bot) \sin{\alpha}\cos{\alpha}  \, ,      \label{mpol} 
 \\ \psi_{zz} & = & \psi_\bot\sin^2{\alpha}+
 \psi_{\|}\cos^2{\alpha}  \, , \nonumber
\end{eqnarray}
where $\psi_{\|}$ is the longitudinal magnetic susceptibility 
(for the external magnetic field along the major axis), and 
$\psi_\bot$ is the transverse one (the field is transverse to 
the major axis of the hole). Formulas for polarizabilities of 
various discontinuities are collected in the handbook 
\cite{AccHandb}, see also Ref.\ \cite{SK2000}.

When the effective dipoles are obtained by substituting beam 
fields (\ref{beamf}) into Eqs.~(\ref{dip}), one can calculate the 
scattered fields as a sum of waveguide eigenmodes excited in the 
chamber by the dipoles, and find the impedance. This approach has 
been carried out for a circular pipe in \cite{SK}, and for an 
arbitrary chamber in \cite{SK92}, for ultrarelativistic beams. 
The fields radiated into the chamber by the effective dipoles 
(\ref{dip}) can be found as a series in TM- and TE-eigenmodes 
\cite{Collin}:
\begin{eqnarray}
\vec{F} = \sum_g\left [ A^{+}_g\vec{F}^{(E)+}_g\theta (z) +   
 A^{-}_g\vec{F}^{(E)-}_g\theta (-z) \right ] +  \label{fexp} \\
 \sum_g\left [ B^{+}_g\vec{F}^{(H)+}_g\theta (z) +   
 B^{-}_g\vec{F}^{(H)-}_g\theta (-z) \right ] \, , \nonumber
\end{eqnarray}
where $\vec F$ means either $\vec E$ or $\vec H$, $g=\{n,m\}$ 
is a 2D index; superscripts '$\pm$' denote waves radiated 
respectively in the positive (+, $z>0$) or negative 
($-$, $z<0$) direction, and $\theta (z)$ is the Heaviside 
step function. The fields $F^{(E)}_g$ of the $g$th TM-eigenmode 
in Eq.~(\ref{fexp}) are expressed \cite{Collin} in terms of 
EFs (\ref{boundpr})
\begin{eqnarray}
E^\mp_z & = & k_g^2 e_g \exp(\pm \Gamma _gz) \, ; 
     \qquad  H^\mp_z = 0 \, ; \nonumber \\
\vec{E}^\mp_t & = & \pm \Gamma _g \vec{\nabla}e_g
        \exp(\pm \Gamma _gz) \, ;               \label{emode} \\
\vec{H}^\mp_t & = & i\omega \varepsilon_0 \hat{z} \times 
        \vec{\nabla} e_g \exp(\pm \Gamma _gz) \, , \nonumber 
\end{eqnarray}
where propagation factors $\Gamma _g=(k_g^2-\omega^2/c^2)^{1/2}$ 
should be replaced by $-i \beta _g$ with 
$\beta _g=(\omega^2/c^2-k_g^2)^{1/2}$ for $\omega/c>k_g$. 
For given values of dipoles (\ref{dip}) the coefficients 
$A^\pm_g$ can be found \cite{SK,SK92} using the 
Lorentz reciprocity theorem 
\begin{equation}
A^{\pm}_g = a_g M_\tau \pm b_g P_\nu \, ,  \label{Apm}
\end{equation}
with 
\begin{equation}
 a_g  = - \frac{i \omega \mu_0}{2 \Gamma _g k_g^2}   
                \nabla_\nu e^h_g \, ;  \quad
 b_g =  \frac{1}{2 \varepsilon_0 k_g^2}  
                \nabla_\nu e^h_g \, .  \label{ab} 
\end{equation}
Here the EFs are taken at the hole location $\vec{b}$ 
on the wall, $e^h_g \equiv e_g(\vec{b}\,)$. 

Similarly, the fields $F^{(H)}_g$ of the TE${}_g$-eigenmode in 
Eq.~(\ref{fexp}) are 
\begin{eqnarray}
H^\mp_z & = & k'^2_g h_g \exp(\pm \Gamma' _gz) \, ; 
     \qquad  E^\mp_z = 0 \, ; \nonumber \\
\vec{H}^\mp_t & = & \pm \Gamma' _g \vec{\nabla}h_g
        \exp(\pm \Gamma' _gz) \, ;             \label{hmode} \\
\vec{E}^\mp_t & = & -i\omega \mu_0 \hat{z} \times 
  \vec{\nabla} h_g \exp(\pm \Gamma' _gz) \, , \nonumber 
\end{eqnarray}
with propagation factors $\Gamma' _g=(k'^2_g-\omega^2/c^2)^{1/2}$ 
replaced by $-i \beta' _g = -i(\omega^2/c^2-k'^2_g)^{1/2}$ when 
$\omega/c>k'_g$. Here EFs $h_g$ satisfy the boundary problem 
(\ref{boundpr}) with the Neumann boundary condition, 
$\nabla_\nu h_g\vert_{\partial S} = 0$, and $k'^2_g$ 
are corresponding eigenvalues, see in Appendix. The TE-mode 
excitation coefficients in the expansion (\ref{fexp}) for the 
radiated fields are 
\begin{equation}
B^{\pm}_g = \pm c_g M_\tau + d_g P_\nu 
                              +  q_g M_z  \, ,  \label{Bpm}
\end{equation}
where 
\begin{eqnarray}
 c_g & = & \frac{1}{2 k'^2_g} \nabla_\tau h^h_g \, ;  
 \quad q_g  =  \frac{1}{2 \Gamma'_g } h^h_g \, ; 
 \nonumber  \\
 d_g & = & - \frac{i\omega }{2 \Gamma'_g 
 k'^2_g} \nabla_\tau h^h_g \, .                 \label{cdq} 
\end{eqnarray}

From Eqs.~(\ref{fexp})-(\ref{cdq}), using the effective dipoles
(\ref{dip}) induced on the discontinuity by the beam fields 
(\ref{beamf}), we find the fields scattered by the 
discontinuity into the vacuum chamber. Note that the dependence
on the beam velocity enters explicitly only in Eq.~(\ref{beamf}).
Now we proceed with the impedance calculation.  

\section{Beam Coupling Impedance of A Small Discontinuity}

\subsection{Longitudinal Impedance}

In a general case, the longitudinal impedance is defined 
(e.g., \cite{SKrev,Z&Kh}) as an integral along the vacuum chamber 
of the (synchronous with the beam) harmonic of the longitudinal 
electric field $E_z$ created in the chamber by the leading charge 
(\ref{chrgt}) divided by the amplitude of the corresponding 
harmonic of the beam current (\ref{chrgt}). The field $E_z$ 
should be taken at the longitudinal position of a unit test 
charge that follows the leading one with the same velocity, 
and should also be integrated over the test-charge transverse 
distribution $t(\vec{r}\,)$. This leads to the formula
\begin{equation}
Z(\omega) = -\frac{1}{q} \int_{-\infty}^{\infty} 
dz \: e^{-i\frac{\omega z}{\beta c}} \int_S d \vec{r}\, 
t(\vec{r}\,) E_z(\vec{r},z;\omega)\, , \label{impdef}
\end{equation}
where normalization $\int_S d\vec{r}\, t(\vec{r}\,)=1$ is assumed. 
Note that the longitudinal field $E_z$ includes both
the beam field (\ref{Ezf}) and the scattered field (\ref{fexp}) 
produced by the discontinuity. The first one is related to the 
space-charge impedance and vanishes at $\gamma \to \infty$; we 
will consider it in Sec.~IV. Here we include into $E_z$ only 
the discontinuity contribution. All $z$-dependence in 
Eq.~(\ref{fexp}) is in the exponents, so integrating over $z$ 
is straightforward. For transverse integration, we expand the 
test-charge transverse distribution $t(\vec{r}\,)$ in EFs as 
$t(\vec{r}\,) = \sum_g t_g e_g(\vec{r}\,)$, where $t_g = \int_S 
 d\vec{r}\, t(\vec{r}\,) e_g(\vec{r}\,)$. The result is
\begin{equation}
Z(\omega) = i\frac{\omega}{q c} \left ( Z_0 M_\tau + 
 \frac{1}{\beta} \frac{P_\nu}{\varepsilon_0} \right )
 \sum_g \frac{t_g\nabla_{\nu}e_g^h}{k^2_g+\kappa^2}\,, \label{imp1}
\end{equation}
where we again use variable $\kappa = \omega/(\beta\gamma c)$ 
introduced earlier in Eq.~(\ref{beamf}).
Substituting the dipole expressions (\ref{dip}), where the beam
fields (\ref{beamf}) are taken at the hole location $(\vec{b},0)$,
leads to 
\begin{eqnarray}
Z(\omega) & = & -i Z_0 \frac{\omega}{c} 
 \frac{\psi_{\tau\tau} - \chi/\beta^2}{2} \times    \label{imp2} 
\\ & \times & \sum_g \frac{f_g \nabla_{\nu}e_g^h}
    {k^2_g+\kappa^2} \sum_g \frac{t_g \nabla_{\nu}e_g^h}
    {k^2_g+\kappa^2} \, .      \nonumber  
\end{eqnarray}
It is convenient to introduce the following notation
\begin{equation}
e_\nu(f;\kappa) \equiv -\sum_g 
 \frac{f_g\nabla_{\nu}e_g^h}{k^2_g+\kappa^2} \, .   \label{enfw}
\end{equation} 
Comparing to Eqs.~(\ref{beamf}), one can see that 
$e_\nu(f;\kappa)$ is related to the transverse harmonics 
of the beam field at the hole location as
\begin{equation}
e_\nu(f;\kappa) = 
 \frac{E_{\nu}(\vec{b},0;\omega)\beta}{Z_0 q}
 = \frac{H_{\tau}(\vec{b},0;\omega)}{q}  \, .     \label{enfw-c}
\end{equation}
As was already mentioned in Sect.~II, the lowest eigenvalue 
$k_g$ is of the order of $1/b$. Therefore, for 
$\omega b/(\beta\gamma c)=\kappa b \ll 1$, the normalized 
transverse field Eq.~(\ref{enfw}) becomes frequency- and 
velocity-independent: 
\begin{equation}
e_\nu(f;0) = -\sum_g f_gk^{-2}_g\nabla_{\nu}e_g^h \,. \label{enf0}
\end{equation} 
The condition $\omega b/(\beta\gamma c) \ll 1$ includes two 
cases: (i) ultrarelativistic limit, $\gamma \to \infty$; and 
(ii) long-wavelength (or low-frequency) limit, when the 
wavelength $\lambda=2\pi/k=2\pi \beta c/\omega $ is large 
compared to the typical cross-section size $b$. Equation 
(\ref{enf0}) has a simple physical interpretation. It gives a
solution for a 2D electrostatic field created on the chamber 
wall in the cross section $S$ by a uniform in $z$ charge, 
equal to $\varepsilon_0$ per unit length of $z$, that has 
the transverse distribution $f(\vec{r}\,)$. From the Gauss 
law, $e_\nu(f)$ satisfies the normalization condition
\begin{equation}
 \oint_{\partial S}\! dl \ e_\nu(f;0) = 1 \, ,  \label{norma}
\end{equation}
where integration goes along the cross-section boundary 
${\partial S}$. For a simple particular case of a circular
cross section with radius $b$, and an axisymmetric charge
distribution $f(\vec{r}\,)=f(r)$ --- it includes an on-axis
point charge, --- Eq.~(\ref{norma}) by itself provides the 
solution, due to the problem symmetry: 
$e_\nu(f;0)=1/(2\pi b)$, cf.\ \cite{SK,SK92}.

When the test charge distribution is identical to that of the
leading charge, $t(\vec{r}\,)=f(\vec{r}\,)$, Eq.~(\ref{imp2}) 
gives the discontinuity impedance for the given transverse 
charge distribution $f(\vec{r}\,)$ of the beam as
\begin{equation}
Z(\omega) = -i Z_0 \frac{\omega}{c} 
 \left [e_\nu(f;\kappa) \right ]^2 
 ( \alpha_m + \frac{\alpha_e}{\beta^2} ) \,,       \label{imp3}
\end{equation}
where we use the effective polarizabilities 
$\alpha_m=\psi_{\tau\tau}/2$ and $\alpha_e=-\chi/2$, 
and $e_\nu(f;\kappa)$ is defined by (\ref{enfw}).

The generalized longitudinal impedance is usually defined 
with the leading and test point charges having different 
transverse displacement from the chamber axis, 
$\vec{s}$ and $\vec{t}$, respectively \cite{SKrev,KGS}. 
In that case, Eq.~(\ref{imp2}) can be rewritten as
\begin{equation}
Z(\vec{s}\,,\vec{t}\,;\omega) = -i Z_0 \frac{\omega}{c} 
 e_\nu(\vec{s}\,;\kappa) e_\nu(\vec{t}\,;\kappa)
 ( \alpha_m + \frac{\alpha_e}{\beta^2} )        \,, \label{imp4}
\end{equation}
where 
\begin{equation}
e_\nu(\vec{r}\,;\kappa) \equiv 
 -\sum_g \frac{e_g(\vec{r}\,) \nabla_{\nu}e_g^h } 
               {k^2_g+\kappa^2} \,.                  \label{enp}
\end{equation}
The last expression follows from Eq.~(\ref{enfw}) because 
$f_g=e_g(\vec{r}\,)$ for a point charge with transverse 
displacement $\vec{r}$ from the chamber axis. 
The impedance (\ref{imp4}) includes higher multipole longitudinal 
impedances. The usual monopole longitudinal impedance is obtained 
from Eq.~(\ref{imp4}) when the charges are on axis, $s \to 0$ 
and $t \to 0$:
\begin{equation}
Z(\omega) = -i Z_0 \frac{\omega}{c} e^2_\nu(0;\kappa)
 ( \alpha_m + \frac{\alpha_e}{\beta^2} )      \,.    \label{imp5}
\end{equation}
In the ultrarelativistic limit, $\beta \to 1$ and $\gamma \to 
 \infty$, the normalized field $e_\nu(0;\kappa)$ becomes 
\begin{equation}
\tilde{e}_\nu \equiv -\sum_g k^{-2}_g e_g(0)
 \nabla_{\nu}e_g^h ,                               \label{enf0p} 
\end{equation}
cf.\ (\ref{enf0}), and Eq.~(\ref{imp5}) coincides with the known 
result for the impedance of a small discontinuity \cite{SK92,KGS}. 

For some particular cases of simple cross sections $S$, one can 
obtain explicit expressions of the normalized field (\ref{enfw}) 
and the longitudinal impedance (\ref{imp5}). The derivation is
presented in Appendices for circular and rectangular cross 
sections. For a circular cross section of radius $b$, 
(\ref{imp5}) takes form
\begin{equation}
Z(\omega) = -i Z_0 \frac{\omega}{c} 
\frac{\alpha_m + \beta^{-2}\alpha_e}{4\pi^2 b^2} 
 \left[ I_0(\kappa b) \right]^{-2} ,               \label{imp5c}
\end{equation}
which coincides, up to notations, with the result in \cite{G&F99}. 
Here $I_0(x)$ is the modified Bessel function of the first kind.

For a rectangular cross section $a \times b$, 
assuming that the hole is located on the side wall at 
$x=a$, $y=y_h$, the longitudinal impedance (\ref{imp5}) is
\begin{eqnarray}
 Z(\omega) & = & -i Z_0 \frac{\omega}{c} 
\frac{\alpha_m + \beta^{-2}\alpha_e}{b^2} \times \nonumber  
 \\  & \times & \left[ \sum_{p=0}^\infty
\frac { (-1)^p \sin [\pi (2p+1)y_h/b ]}
 {\cosh (\pi u_{2p+1}/2)} \right]^2             \ , \label{imp5r}
\end{eqnarray}
where $u_m=a\sqrt{m^2/b^2+\kappa^2/\pi^2}$.
In the ultra\-re\-la\-tivistic limit, Eq.~(\ref{imp5r}) 
coincides with the result in \cite{KGS}, 
see Appendix~B for detail.

For the case of an axisymmetric small obstacle on the wall of 
a circular beam pipe --- like a small enlargement (cavity) or 
an iris --- the longitudinal impedance turns out to be very 
similar to Eq.~(\ref{imp5}):
\begin{eqnarray}
Z(\omega) & = & -i Z_0 \frac{\omega}{c} 
 e^2_\nu(0;\kappa) \, 2\pi b \, ( \tilde{\alpha}_m 
        + \frac{\tilde{\alpha}_e}{\beta^2} )        \label{imp5a}
\\ & = & -i Z_0 \frac{\omega}{c} 
\frac{\tilde{\alpha}_m + \beta^{-2}\tilde{\alpha}_e}{2\pi b} 
\left[ I_0(\kappa b) \right]^{-2}         \nonumber , 
\end{eqnarray}
where the effective polarizabilities $\tilde{\alpha}_m$ and 
$\tilde{\alpha}_m$ are now defined per unit length of the 
circumference $2\pi b$ of the chamber cross section (circle) $S$.
In transition from the first line in Eq.~(\ref{imp5a}) to the
second one, we used the known expression for 
$e_\nu(0;\kappa)$ in a circular cross section, 
cf.\ Eq.~(\ref{imp5c}). Essentially, for an axisymmetric 
discontinuity $\alpha_{m,e}$ in Eqs.~(\ref{imp5}) and 
(\ref{imp5c}) are replaced by $2\pi b\, \tilde{\alpha}_{m,e}$.
In the ultrarelativistic limit, $\beta \to 1$, Eq.~(\ref{imp5a}) 
coincides with the previous results \cite{K&S,SK97}.
It is worth mentioning that if an axisymmetric enlargement 
has area $A$ of the longitudinal cross section, its magnetic 
polarizability is $\tilde{\alpha}_m=A$, while for an 
axisymmetric protrusion (iris) of the same cross-section 
area $\tilde{\alpha}_m=-A$.
The electric polarizability $\tilde{\alpha}_e$ can be found by 
solving a 2D electrostatic problem, see examples in 
\cite{K&S,SK97,SK2000}. It is positive for protrusions and
negative for enlargements, so that in both cases 
$\tilde{\alpha}_m$ and $\tilde{\alpha}_e$ have opposite signs.  

The longitudinal impedances (\ref{imp4}), (\ref{imp5}), and 
also (\ref{imp5a}), depend on the beam velocity in two ways: 
via $e_\nu(\vec{s}\,;\kappa)$ and in the combination 
of polarizabilities $(\alpha_m + \alpha_e/\beta^2)$.
The first dependence enters via the parameter $\kappa b =
\omega b/(\beta \gamma c)$, as one can see from Eqs.~(\ref{enp}), 
(\ref{imp5c})-(\ref{imp5a}). For $\kappa b \ll 1$ the factor 
$e^2_\nu(\vec{s}\,;\kappa)$ is close to its 
ultra\-relativistic limit, while at $\kappa b > 1$ it 
decreases exponentially to zero. We should emphasize that
for $\beta < 1$ the monopole longitudinal impedance 
depends on the beam position in the chamber cross section,
unlike its ultra\-relativistic counterpart \cite{Z&Kh}. 
For a circular cross section this dependence takes a 
particularly simple form as an additional factor of 
$I^2_0(\kappa t)$ in (\ref{imp5c}) and (\ref{imp5a}), 
where $t$ is the beam transverse displacement from the 
chamber axis, see Eq.~(\ref{enpc}) in the Appendix. 

In the combination $\alpha_m + \alpha_e/\beta^2$ the electric 
contribution is enhanced as the beam velocity decreases. 
This sum (more exactly, this difference, because $\alpha_m$ 
and $\alpha_e$ always have opposite signs) can either vanish
for some values of $\beta$, or become much larger than its
ultra\-relativistic limit $\alpha_m + \alpha_e$. It vanishes
when $0 < \beta=\sqrt{-\alpha_e/\alpha_m} <1$. We should note 
that such situation occurs only for discontinuities like holes 
or chamber enlargements (small cavities), since for them
$\alpha_m > |\alpha_e|$ \cite{SK92}. For instance, a circular 
hole of radius $h$ in a thin wall has $\alpha_m=4h^3/3$ and 
$\alpha_e=-2h^3/3$, so that $\sqrt{-\alpha_e/\alpha_m} =
1/\sqrt{2}$. The impedance (\ref{imp5}) of the hole, which 
is inductive for relativistic beams, changes its sign for 
$\beta < 1/\sqrt{2}$ becoming a ``negative inductance''.
On the other hand, for protrusions and irises the impedance
remains inductive for any beam velocity because of
$\alpha_e > |\alpha_m|$, cf.\ \cite{SK97,K&S}. As a simple 
example, a semi-spherical protrusion (bump) of radius 
$a$ on the wall has polarizabilities $\alpha_e=2\pi a^3$ 
and $\alpha_m=-\pi a^3$ \cite{SK97}. 

For all small discontinuities the impedance vanishes at very 
slow beam velocities, when $\beta \to 0$, since the fast 
decrease of the factor $e^2_\nu(\vec{s}\,;\kappa)$ 
suppresses the growth due to $\alpha_e/\beta^2$. This
behavior is illustrated in Fig.~1 for a hole and in 
Fig.~2 for a protrusion. The impedance magnitude can exceed 
the ultra\-relativistic value many times. In fact, the 
ratio $Z(\beta)/Z(1)$ for $\omega b/c=0.1$ in Fig.~1 reaches 
-83.3 at $\beta=0.062$, and in Fig.~2 its maximum is 167.5, 
well outside the shown range. The extremes become even larger 
for lower frequencies.


\begin{figure}[htb]
\resizebox{8.5cm}{!}
 {\includegraphics{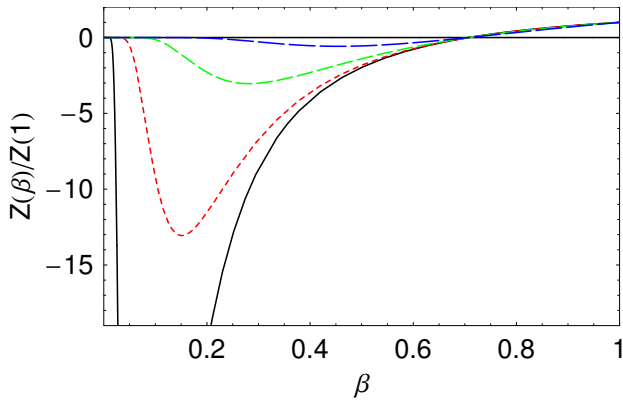}}
\caption{The ratio of the longitudinal impedance (\ref{imp5c})
to its relativistic value for a circular hole in a round pipe 
versus $\beta=v/c$ for $\omega b/c=0.1,0.25,0.5,1$ 
(solid, short-dashed, dashed, and long-dashed curves).}
\end{figure}

\begin{figure}[htb]
\resizebox{8.5cm}{!}
 {\includegraphics{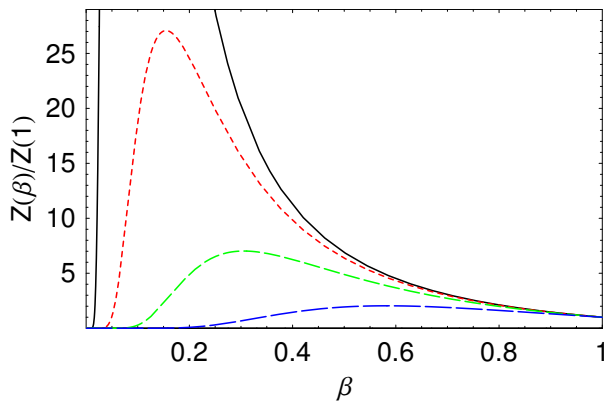}}
\caption{The same, for a small semi-spherical protrusion.}
\end{figure}

Another interesting example of $\beta$-dependence is for 
long elliptic slots parallel to the chamber axis. If the 
ellipse semi-axes $w,l$ satisfy $w \ll l \ll b$, the
leading terms ($\propto w^2l$) of $\alpha_m$ and $\alpha_e$ 
for a thin wall cancel each other \cite{SK,SK92} in the 
ultra\-relativistic limit:
$$ \alpha_m + \alpha_e \approx \frac{\pi w^4}{3l}
 ( \ln{\frac{4l}{w}} - 1 ) \, .
$$
For $\beta<1$ there is no such cancellation, so that
$$ \alpha_m + \frac{\alpha_e}{\beta^2} \approx 
 -\frac{\pi w^2 l}{3 \beta^2 \gamma^2} + 
 \frac{\pi w^4 }{3 l \beta^2} ( \frac{1+\beta^2}{2} 
 \ln{\frac{4l}{w}}- \frac{1}{4}- \frac{3 \beta^2}{4} ),
$$
which for small $\beta$ has an opposite sign and larger 
magnitude compared to its relativistic limit.

The longitudinal impedance in Eqs.~(\ref{imp3})-(\ref{imp5})
is purely inductive. In the sense of a perturbative expansion 
in the small parameters $|\alpha_m|/b^3 \ll 1$ and 
$|\alpha_e|/b^3 \ll 1$, it is the first-order term.
In Ref.\ \cite{KGS} higher-order terms were studied 
by taking into account radiative corrections to the fields
near the discontinuity. The second-order term has both an 
imaginary and real part; the last one appears 
only at frequencies above the chamber cutoff. 
The real part of the impedance includes contributions from 
only a finite number of the eigenmodes propagating in the 
chamber at a given frequency, i.e.\ those with 
$k_g<\omega /c$ or $k'_g<\omega /c$. The dependence of 
$Re\,Z$ on frequency is complicated: it has sharp peaks near 
the cutoffs of all propagating eigenmodes of the chamber, 
but increases on average with the frequency increase.
Well above the chamber cutoff, i.e.\ when $\omega b/c \gg 1$ 
(still $\omega h/c \ll 1$ to justify the Bethe approach), this 
averaged dependence can be derived in exactly the same way 
as in \cite{KGS} for ultrarelativistic beams. The result is  
\begin{equation}
 Re\,Z(\omega) = \frac{Z_0 \omega^4}{3 \pi c^4} 
 e^2_\nu(0;\kappa) ( \alpha^2_m + \alpha^2_{ms} + 
 \frac{\alpha^2_e}{\beta^2} )         \, ,        \label{ReZhf}
\end{equation}
where $\alpha_{ms} \equiv \psi_{\tau z}/2$, cf.\ 
Eqs.~(\ref{dip})-(\ref{mpol}). Alternatively, the same answer 
can be obtained by calculating the energy radiated by the 
effective dipoles (\ref{dip}) into a half-space. The physical 
reason for this coincidence is that at frequencies well above 
the cutoff  they radiate into the waveguide the same energy as 
into an open half-space.

\subsection{Transverse Impedance}

Since we are interested in the dipole transverse 
impedance, it is convenient to consider the excitation by a
dipole: two point charges with the opposite signs displaced 
from the chamber axis by $\vec{s}$ and $-\vec{s}$, 
correspondingly, instead of Eq.~(\ref{chrgt}), moving at 
velocity $\beta c$ along the axis. Such a change would 
modify the expression (\ref{imp4}) for the longitudinal 
impedance replacing $e_\nu(\vec{s}\,;\kappa)$ by
$e^{dip}_\nu(0;\kappa) = \vec s \cdot 
\vec{\nabla} e_\nu(0;\kappa)$, which is the second 
(dipole) term of its Taylor expansion, with 
$e_\nu(0;\kappa)$ being the first (monopole) one.
After normalizing the excitation to the unit 
transverse beam displacement (dividing by $s$),  
we integrate the synchronous harmonic of the transverse force 
acting on a test charge displaced from the axis by $\vec{t}$, 
and then divide by the harmonic amplitude of the dipole 
moment creating the deflecting force: 
\begin{eqnarray}
\vec Z_\bot(\omega) & = & -\frac{i}{q s} 
 \int_{-\infty}^{\infty} dz \: \exp 
 \left(-i\frac{\omega z}{\beta c}\right) \times  \label{imptdef}
\\ && \qquad \times \left [ \vec{E}_\bot(\vec{t}\,,z;\omega) 
 + Z_0 \beta \hat{z} \times \vec{H}_\bot(\vec{t}\,,z;\omega) 
                          \right ] \, .         \nonumber 
\end{eqnarray}
where the usual definition of the dipole transverse impedance
assumes the limit of $t \to s \to 0$, e.g.\ \cite{Z&Kh} or 
\cite{SKrev}. The integration result includes contributions 
from both TM- and TE-modes excited by the effective dipoles 
in the beam pipe, unlike the expression for the longitudinal 
impedance (\ref{imp4}), where only TM-modes contribute. 

There is an alternative way to find the transverse impedance.
Since we already have expression (\ref{imp4}) for the 
generalized longitudinal impedance $Z(\vec s,\vec t\:;\omega)$, 
we can apply the Panofsky-Wenzel theorem, e.g., in \cite{Z&Kh}. 
According to the theorem, the transverse impedance can be 
derived as 
\begin{equation}
\vec Z_\bot(\vec s,\vec t\:;\omega)=\frac{\beta c}{\omega s} 
 \, \vec {\nabla}_t Z(\vec s,\vec t\:;\omega) \ ,     \label{PW}
\end{equation}
where the longitudinal impedance is calculated with the 
dipole excitation as discussed above. 
If we follow this way, which is simpler in our case, 
the result will include only TM-mode contributions, or, 
in other words, the EFs of (\ref{boundpr}), 
since only they enter Eq.~(\ref{imp4}) for the longitudinal
impedance. We used the relations between the transverse 
components of the beam fields, cf.\ Eq.\ (\ref{beamf}), 
to prove that the two results obtained are equivalent. 
One should remark that definition (\ref{imptdef}) sometimes
includes an extra factor $1/\beta$, e.g.\ in \cite{SKrev}.
For consistency with the relation (\ref{PW}), here we use 
the transverse impedance definition without that extra factor, 
i.e., Eq.~(\ref{imptdef}). 

The final expression for the transverse dipole impedance of a 
discontinuity is
\begin{equation}
\vec Z_\bot(\omega) = -i Z_0 \beta
 ( \alpha_m + \frac{\alpha_e}{\beta^2} )
 [ \hat{s} \cdot \vec{d}\,(\kappa) ] \,
 \vec{d}\,(\kappa)                      , \label{impt}
\end{equation}
where $\hat{s}=\vec{s}/s$ is a unit vector in the direction
of the beam deflection from the chamber axis, and 
\begin{equation}
 \vec{d}\,(\kappa) \equiv 
 \vec{\nabla} e_\nu(0;\kappa) = -\sum_g 
 \frac{ \vec{\nabla} e_g(0) \nabla_{\nu}e_g^h} 
                          {k^2_g+ \kappa^2}   \, . \label{gren0}
\end{equation}
The impedance dependence on the discontinuity shape is obviously
the same as for the longitudinal impedance. The direction of the 
vector of the transverse impedance (\ref{impt}) gives the 
direction of the deflecting force acting on a displaced beam. 
As one can see from Eq.~(\ref{impt}), this direction is defined 
by vector $\vec{d}$, Eq.~(\ref{gren0}), while the force 
magnitude varies depending on the relative direction of the beam 
displacement $\vec{s}$ with respect to $\vec{d}$ via the scalar 
product $\hat{s} \cdot \vec{d}$.

For a circular beam pipe, the transverse impedance (\ref{impt})
can be written, making use of Eq.~(\ref{gren0c}) of Appendix, 
simply as
\begin{equation}
\vec Z_\bot(\omega) = -i Z_0 \beta
 \frac{\alpha_m + \beta^{-2} \alpha_e}{\pi^2 b^4}
 \left [ \frac{\kappa b}{2 I_1(\kappa b)} \right ]^2
 \! (\hat{s} \cdot \hat{h}) \ \hat{h} \, ,         \label{imptc}
\end{equation}
where $\hat{h}=\vec{b}/b$ is a unit vector in the chamber 
cross section $S$ directed from the axis to the hole 
(discontinuity). In this case, one can rewrite the dot 
product in a more conventional form, 
$\hat{s} \cdot \hat{h} = \cos (\varphi_s-\varphi_h)$,
where $\varphi_s$ is the azimuthal angle of the beam position
in the cross-section plane, and $\varphi_h$ is the azimuthal 
angle in the direction to the hole. Therefore, the deflecting 
force in the circular pipe is directed to (or opposite to) the 
discontinuity, and its magnitude depends on the angle between 
the hole and the beam transverse displacement, as was pointed out 
already in the relativistic case \cite{SK}. Equation 
(\ref{imptc}) agrees, up to notations, with the result obtained 
in \cite{G&F99}. 

In a general case, it is sometimes convenient to rewrite the 
dipole transverse impedance (\ref{impt}) as
\begin{equation}
 \vec Z_\bot(\omega) = -iZ_0 \beta
 (\alpha_m + \frac{\alpha_e} {\beta^2}) \, d^2(\kappa) \, 
 \hat{d} \cos (\varphi_s - \varphi_d)  \, .            \label{Zt} 
\end{equation}
In particular, this form is more convenient for the rectangular
chamber. Here $d=\sqrt{d^2_x+d^2_y}$, where $x,y$ are the 
horizontal and vertical coordinates in the chamber cross section 
$S$, and $d_x$ and $d_y$ are projections of $\vec{d}$, cf.\ 
Appendix~B; $\varphi _s =\varphi _t$ is again the azimuthal 
angle of the beam position in the cross-section plane; $\hat{d} =
\vec{d}/d$ is a unit vector in this plane in the direction 
of $\vec{d}$, and $\varphi_d$ is the corresponding azimuthal angle. 
Obviously, $d_x=d \cos \varphi_d$ and $d_y=d \sin \varphi_d$.
It is seen from Eq.~(\ref{Zt}) that the angle $\varphi _d$ shows 
the direction of the transverse-impedance vector $\vec Z_\bot$ and, 
therefore, of the beam-deflecting force. Moreover, the magnitude of 
$Z_\bot$ is maximal when the beam is deflected along this direction 
and vanishes when the beam offset is perpendicular to it. For a 
circular pipe, $\varphi_d = \varphi_h$, as was discussed above.
For a general cross section, this is not the case even in the
relativistic limit; see \cite{SK92} for rectangular and elliptic 
chambers. As an illustration, in Fig.~3 we plot the relation 
between $\varphi_d$ and $\varphi_h$ for a square chamber for 
a few different values of parameter $\kappa b = 
\omega b/(\beta \gamma c)$. The case $\kappa b = 0$ corresponds
to the relativistic (or low-frequency) limit; as $\kappa b$ 
increases, the difference between $\varphi_d$ and $\varphi_h$
becomes smaller.


\begin{figure}[htb]
\resizebox{8.5cm}{!}
 {\includegraphics{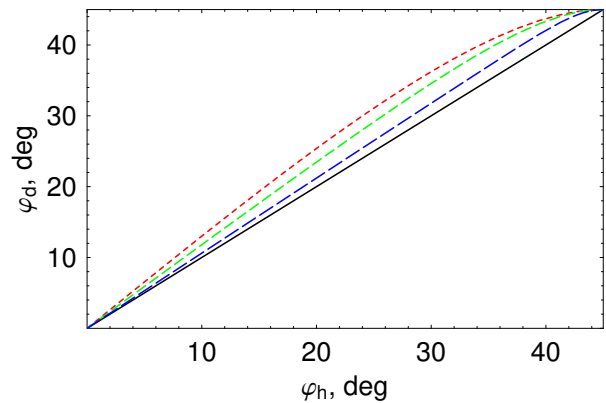}}
\caption{Direction $\varphi_d$ of the transverse impedance 
(\ref{Zt}) versus hole position on the side wall of a square 
chamber for $\omega b/(\pi \beta \gamma c)=0,3,10$ (short-dashed, 
dashed, and long-dashed curves). The solid line is for a 
circular pipe.}
\end{figure}

\begin{figure}[htb]
\resizebox{8.5cm}{!}
 {\includegraphics{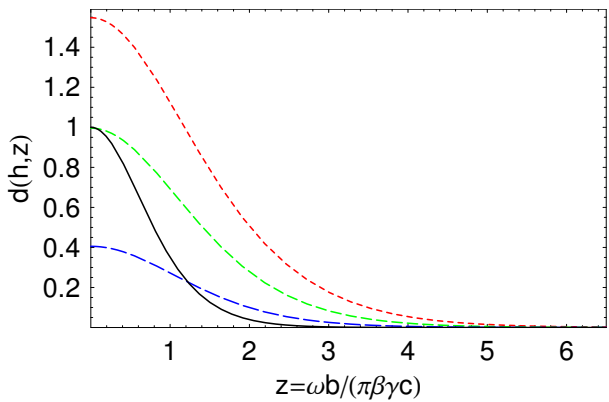}}
\caption{Normalized magnitude of function $d$ in Eq.~(\ref{Zt}) 
versus $\kappa b/\pi$ for three hole positions 
$h=y_h/b=0.5,0.75,0.9$ on the side wall of a square chamber 
(short-dashed, dashed, and long-dashed curves). The solid 
line shows for comparison $\kappa b/2/I_1(\kappa b)$ from 
Eq.~(\ref{imptc}).}
\end{figure}

In Fig.~4, the behavior of $d(\kappa)$ in Eq.~(\ref{Zt})
is illustrated for a square cross section of the vacuum chamber.
This function is given by Eq.~(\ref{grenpr}) in the Appendix; 
for plotting $d(\kappa)$ is multiplied by $a b$ to make 
it dimensionless. For comparison, the same plot shows a similar
dependence for the circular cross section, see Eq.~(\ref{imptc}).
Figure~4 demonstrates also the impedance dependence on the 
position of the discontinuity on the chamber wall: 
$y_h/b=0.5$ corresponds to the middle of the wall, 0.75 
is one quarter from the corner, and 0.9 is close to the corner. 
Figures 3-4, as well as Figs.~1-2, were produced using 
{\it Mathematica} \cite{Math}. It takes less than ten terms
in the series (\ref{grenpr}) and (\ref{enpr0}) to get a very
accurate answer for the sum, which is not surprising since the 
10th term in (\ref{enpr0}) is of the order of $O(e^{-30})=
O(10^{-13})$. However, even with hundreds of terms 
{\it Mathematica} gives the results almost as quickly as those 
involving modified Bessel functions for the circular pipe in 
Figs.~1-2.
 
One should remind that the impedance magnitude decreases even 
faster with the frequency increase for a fixed $\beta$ than 
shown in Fig.~4, since $Z_\bot \propto d^2$. The full 
$\beta$-dependence of the transverse impedance includes, 
of course, the shape factor $\beta \alpha_m + \alpha_e/\beta$; 
overall, it is similar to that of the longitudinal impedance
illustrated in Figs.~1-2.

\section{Space-Charge Impedance}

We will show in this Section that the same approach as above --- 
based on field expansions in cross-section eigenfunctions --- 
works equally well for calculating the space-charge impedance of a 
uniform waveguide with an arbitrary simply-connected cross section. 
While this subject seems to be somewhat off topic for the present 
paper, the result comes as a by-product of the considerations 
in the previous Sections. The synchronous harmonic of the 
longitudinal beam field produced in a uniform beam pipe with 
the cross section $S$ by the current (\ref{currw}) is given by 
Eq.~(\ref{Ezf}). Using the impedance definition (\ref{impdef}) with
the finite integration length $L$ and assuming the same transverse
charge distribution for the test charge as for the source, 
$t(\vec{r}\,)=f(\vec{r}\,)$, we obtain the space-charge 
longitudinal impedance per unit length of the chamber 
\begin{equation}
 \frac{Z^{sc}(\omega)}{L} = i \, \frac{\omega}{c} 
 \frac{Z_0}{\beta^2 \gamma^2} 
    \sum_g \frac{f^2_g}{k^2_g + \kappa^2}  \       \label{Zsc}      
\end{equation}
for a beam with the transverse charge distribution $f(\vec{r}\,)$.
The space-charge impedance (\ref{Zsc}) is expressed in terms of 
eigenvalues $k_g$ of the boundary problem (\ref{boundpr}) and 
the expansion coefficients $f_g$ of the charge distribution in 
EFs (\ref{boundpr}), $f_g=\int_S d\vec{r}\, f(\vec{r}\,) 
e_g(\vec{r}\,)$. Every term of the series in Eq.~(\ref{Zsc})
is positive, so that the space-charge impedance always remains 
a ``negative inductance''. It is important to emphasize that 
the dependence on the beam charge distribution is an 
essential feature of the space-charge impedance. For example,
one can not use a pencil beam (point charge) in (\ref{Zsc}): 
the space-charge impedance diverges as the beam transverse 
size vanishes, e.g.\ \cite{Z&Kh}. 
On the contrary, the geometrical impedances discussed in the 
previous sections are generally independent of the beam 
properties, at least in the relativistic limit, and for that 
reason they are usually calculated in the simplest way, that 
is with an on-axis pencil beam.   

Let us compare Eq.~(\ref{Zsc}) with the  
conventional form of the space-charge impedance in the 
long-wavelength approximation, $\kappa b \ll 1$,
\begin{equation}
 \frac{Z^{sc}(\omega)}{L} = i \, \frac{\omega}{c} 
 \frac{Z_0}{\beta^2 \gamma^2} \frac{1}{2\pi} g_L \ , \label{Zsc0}      
\end{equation}
where $g_L$ is called the longitudinal $g$-factor, 
e.g.\ \cite{Z&Kh,SK99}. One can conclude that the sum in 
(\ref{Zsc}) is proportional to the $g$-factor:
\begin{equation}
 S(f;\kappa) \equiv \sum_g \frac{f^2_g}{k^2_g + \kappa^2} 
                   = \frac{1}{2\pi} g_L(\kappa) \ ,   \label{sig}      
\end{equation}
which now depends on frequency and beam velocity, in addition 
to its familiar dependence on the beam charge distribution. 
The usual value of $g$-factor is obtained as the limit at
$\kappa \to 0$, i.e.\ $g_L=g_L(0)$. Using a term-by-term 
consideration of series (\ref{sig}), it is easy to prove that 
$g$-factor is positive, $g_L(\kappa) > 0$, and that frequency 
corrections reduce its value: $g_L(\kappa) < g_L(0)$.

In the well-studied particular case of a circular cross section, 
we consider the beam in the form of a thin axisymmetric ring of 
radius $a$, $f(\vec{r}\,) = \delta(r-a)/(2\pi a)$. The summation 
is performed in Appendix~A, and the result
\begin{equation} 
 S(f;\kappa) = \frac{I_0(\kappa a)}{2 \pi} 
 \left[ K_0(\kappa a) - I_0(\kappa a) 
 \frac{K_0(\kappa b)}{I_0(\kappa b)} \right]      \label{sigc}
\end{equation}  
coincides with the familiar one, e.g.\ in \cite{RLG2000}. 
In the long-wavelength limit $\kappa b \ll 1$, Eq.~(\ref{sigc})
gives the well-known $g$-factor of a hollow beam
$S = \ln{(b/a)}/(2\pi)$. 

For a rectangular chamber of width $a$ and height $b$, 
to simplify calculations, we choose somewhat exotic beam charge 
distribution: a centered hollow beam with a similar rectangular 
profile $\zeta a$, $\zeta b$, where the beam-size scale
$0<\zeta<1$. Even in a simpler case of a square chamber, $a=b$, 
the space-charge factor (\ref{sig}) of the square hollow beam 
$\zeta a \times \zeta a$ is given by a rather long expression 
(see Appendix~B): 
\begin{eqnarray}
 S(f;\kappa) & = & \frac{2}{\pi^3 \zeta^2} \sum^\infty_{p=0} 
 \frac{\sin [\pi(2p+1)\zeta/2]}{2p+1} \times        \nonumber  
\\ && \qquad \times \frac{\sinh [\pi u_{2p+1}(1-\zeta)/2]}
 {u_{2p+1}  \cosh (\pi u_{2p+1} /2)}  \times        \label{sigr} 
\\ & \times & \left \{ \frac{\sin[\pi(2p+1)\zeta/2]}{2p+1} 
           \cosh (\pi u_{2p+1} \zeta /2) + \right.  \nonumber 
\\ && \quad \left. \frac{\cos[\pi(2p+1)\zeta/2]}{u_{2p+1}}
     \sinh (\pi u_{2p+1} \zeta /2) \right \} ,      \nonumber     
\end{eqnarray} 
where $u_n=\sqrt{n^2+(\kappa a/\pi)^2}$. 
We found analytically that in the relativistic (or 
long-wavelength) limit, $\kappa a \to 0$, the leading term 
of Eq.~(\ref{sigr}) is $\ln{(1/\zeta)}/(2\pi)$, exactly 
the same as in the case of a circular pipe above. 
The series (\ref{sigr}) converges slower than 
the series in (\ref{imp5r}) or (\ref{grenpr}), 
so up to 2500 terms were included in our {\it Mathematica} 
computations for this series. Still it took only a couple 
of minutes of CPU time on a PC to produce Figs.~5-6. 
They illustrate the behavior of $S(f;\kappa)$ 
(\ref{sigr}) as a function of the frequency and beam size.
It is interesting that the sum $S(f,0)$ (relativistic 
or static limit) is very close to its $\ln$ asymptotic 
even for the beam size as large as one half of the pipe 
size, $\zeta=0.5$, cf.\ Fig.~5. The frequency dependence 
of the space-charge impedance (more precisely, of the 
$g$-factor) plotted in Fig.~6 is weaker than that for 
the impedances of small discontinuities in Fig.~4.


\begin{figure}[htb]
\resizebox{8.5cm}{!}
 {\includegraphics{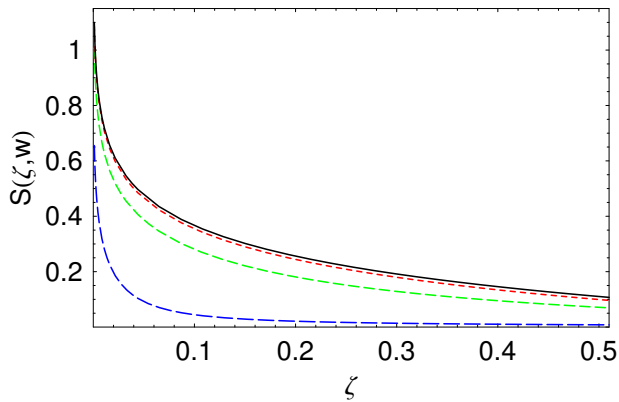}}
\caption{Function $S$ (\ref{sigr}) for a square chamber 
versus the ratio $\zeta$ of the beam size to the chamber size
$a$ at three different values of $w=\kappa a/\pi=0,1,10$
(short-dashed, dashed, and long-dashed curves). The solid 
curve shows the analytical relativistic limit 
$\ln(1/\zeta)/(2\pi)$.}
\end{figure}

\begin{figure}[htb]
\resizebox{8.5cm}{!}
 {\includegraphics{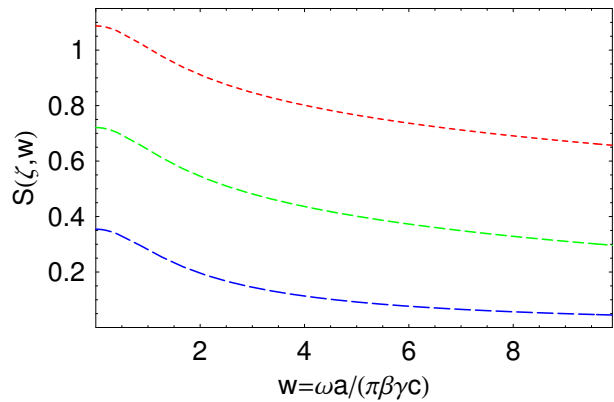}}
\caption{The same versus $w=\kappa a/\pi$ for 
three different beam sizes $\zeta a$, $\zeta=0.001,0.01,0.1$
(short-dashed, dashed, and long-dashed curves).}
\end{figure}

\section{Discussion}

The approach of Refs.~\cite{SK92,KGS}, where the impedances 
of small discontinuities were calculated in the 
ultrarelativistic case, was extended to beams with
an arbitrary velocity and transverse charge distribution.
The analytical approach presented above provides a general 
picture of the coupling impedances for small discontinuities
of the vacuum chamber with an arbitrary cross section in a 
wide frequency range, up to frequencies well above the cutoff. 
The upper limit on the frequency is imposed by the applicability 
of the Bethe theory: the wavelength must be large compared 
to the typical size of the discontinuity. 

We concentrated mostly on the leading (imaginary) part of 
the impedances created by discontinuities. In a general 
case, it was shown that the coupling impedances of small 
discontinuities depend on the beam velocity in two ways: 
first, through the combination of polarizabilities 
$(\alpha_m+\alpha_e/\beta^2)$, and second, via the parameter 
$\kappa = \omega /(\beta \gamma c)$ that enters into the 
factors $e_\nu(\kappa)$ for the longitudinal impedance,
see Eqs.~(\ref{imp5c})-(\ref{imp5a}), and $d(\kappa)$ for 
the transverse one, Eqs.~(\ref{impt})-(\ref{Zt}). 
The first factor contains all the impedance dependence 
on the discontinuity shape, while the second one accounts 
for effects of the chamber cross section and 
the discontinuity location on the cross-section boundary. 
The second factor behaves in the same way for 
various cross sections: it decreases from its value at 
$\kappa = 0 $, which can be interpreted as either the 
ultra\-relativistic or long-wave\-length limit, as $\kappa$
increases. The decrease rate depends to some extent on 
the cross-section shape and the discontinuity position, but
for $\kappa b > 1$ the decrease is fast (exponential), as 
evidenced by Eqs.~(\ref{imp5c}), (\ref{imp5a}) and 
(\ref{imptc}), as well as by Fig.~4. The overall impedance 
dependence on the beam velocity for small discontinuities 
was discussed in Sec.~III. The most interesting feature is 
that the impedance magnitude for some $\beta <1$ can exceed 
its relativistic value many times, see Figs.~1-2.

The impedances for the circular and rectangular cross 
sections of the vacuum chamber were derived from the 
obtained formulas by substituting corresponding EFs. 
The expressions can be useful for calculating the beam 
coupling impedances in particle accelerators with 
non-ultra\-relativistic beams. For the circular pipe 
our results agree with those obtained earlier for $\beta <1$
in \cite{RLG2000,G&F99}. In a similar way, the results for 
an elliptic chamber can be expressed in terms of Mathieu 
functions; it seems unlikely, however, that such a series 
would be convenient for calculations. 
For a complicated cross-section shape the relativistic-limit 
factors $e_\nu(0)$ and $d(0)$ can be found numerically by 
solving a 2D electrostatic problem, see \cite{SK92} 
for an elliptical pipe. Applying the results obtained above, 
one can give then an impedance estimate for a given beam
velocity in such a chamber. 

We also demonstrated that the same technique --- using 
the field expansion into a series of cross-section 
eigenfunctions -- works well for calculating the 
space-charge impedance of a uniform vacuum chamber 
with an arbitrary cross section. A generalized 
expression $g(\kappa)$ for the longitudinal space-charge 
$g$-factor is derived. It depends on the beam velocity and 
on frequency, in addition to the usual dependence on the 
transverse beam-charge distribution. The generalized 
$g$-factor $g(\kappa)$ decreases monotonically as $\kappa$ 
increases: $g(0)\ge g(\kappa)>0$. It is worthwhile to notice 
that the long-wave\-length (or relativistic) limiting value 
$g(0)$ can be calculated numerically even for very 
complicated vacuum chambers, e.g., those with screening 
wires and ceramic insertions \cite{SK99}.

\appendix

\section{Circular Chamber}

For a circular cross section of radius $b$ the eigenvalues 
$k_{nm}=\mu_{nm}/b$, where $\mu_{nm}$ is $m$th zero of the 
Bessel function $J_n(x)$, $n = 0,1,2, \ldots$, and 
$m = 1,2, \ldots$. The normalized EFs in circular co-ordinates
$(r,\varphi)$ are 
\begin{equation} 
e_{nm}(r,\varphi) = \frac{J_n(k_{nm}r)}{\sqrt{N^E_{nm}}} 
    \left \{ \begin{array}{c} 
      \cos {n\varphi} \\ \sin {n\varphi} \end{array} 
    \right \}   \, ,                          \label{ecnm}
\end{equation} 
with $N^E_{nm} = \pi b^2 \epsilon_n J^2_{n+1}(\mu_{nm})/2$, 
where $\epsilon_0=2$ and $\epsilon_n=1$ for $n \ne 0$.
For TE-modes, $k'_{nm}=\mu'_{nm}/b$ with $J'_n(\mu'_{nm})=0$, 
and 
\begin{equation} 
h_{nm}(r,\varphi) = \frac{J_n(k'_{nm}r)}{\sqrt{N^H_{nm}}} 
    \left \{ \begin{array}{c} 
      \cos {n\varphi} \\ \sin {n\varphi} \end{array} 
    \right \}   \, ,                          \label{hcnm}
\end{equation} 
where $N^H_{nm} = \pi b^2 \epsilon_n (1-n^2/\mu'^2_{nm})
J^2_n(\mu'_{nm})/2$. 

The series (\ref{enp}) for the circular cross section can be 
easily summed
\begin{eqnarray}
e_\nu(\vec{r}\,;\kappa) &=& \frac{2}{\pi b}
 \sum^\infty_{n=0}\frac{\cos\left[n(\varphi-\varphi_h)\right]}
 {\epsilon_n} \times          \nonumber \\
 & \times & \sum^\infty_{m=1} 
 \frac{\mu_{nm}J_n(\mu_{nm}r/b)}{J_{n+1}(\mu_{nm})
 \left[ \mu^2_{nm}+(\kappa b)^2 \right]}   \nonumber
\\ &=& \frac{1}{\pi b} \sum^\infty_{n=0}
 \frac{\cos\left[n(\varphi-\varphi_h)\right]}{\epsilon_n} 
 \frac{I_n(\kappa r)}{I_n(\kappa b)} ,              \label{enpc}
\end{eqnarray}
where $I_n(x)$ are the modified Bessel functions, and
$\kappa = \omega/(\beta \gamma c)$.
For the last step in the above equation the summation was 
performed using formulas from \cite{BEII}. The impedances 
of a hole in a circular pipe for $\beta<1$ were derived in 
\cite{Palumbo96} in terms of sums similar to that in the second 
line of Eq.~(\ref{enpc}). The resulting sum (\ref{enpc}) is 
essentially a multipole expansion. 
In the limit of an on-axis beam, $r \to 0$, it becomes simply 
\begin{equation} 
e_\nu(0;\kappa) = \frac{1}{2\pi b} 
 \frac{1}{I_0 (\kappa b)}                  \, .  \label{enpc0}
\end{equation} 
The result simplifies even further in the relativistic limit. 
For $\gamma \to \infty$, $e_\nu(0;\kappa) \to e_\nu(0;0) \equiv 
\tilde{e}_\nu$ and becomes $\tilde{e}_\nu = 1/(2\pi b)$, 
which also follows from the Gauss law. Then the inductive 
impedance (\ref{imp5}) takes an especially simple form 
derived earlier in Refs.~\cite{SK,RLG}. 

For calculating the transverse impedance, we need to find 
the gradient of the field (\ref{enpc}) at the origin, cf.\ 
definition (\ref{gren0}). Only the second term ($n=1$) in 
the series (\ref{enpc}) gives a non-vanishing contribution 
when $r \to 0$:
\begin{equation}
 \vec{d}\,(\kappa) = 
 \vec{\nabla} e_\nu(0;\kappa) = \frac{1}{\pi b^2}  
 \frac{\kappa b}{2 I_1(\kappa b)} \, \hat{h} \,,  \label{gren0c}
\end{equation}
where $\hat{h}=\vec{b}/b$ is a unit vector in the chamber 
cross section $S$ directed from the axis to the hole 
(discontinuity). In the limit of $\gamma \to \infty$, 
Eq.~(\ref{gren0c}) simplifies to $\vec{d}=\hat{h}/(\pi b^2)$,
in agreement with \cite{SK,RLG}.

For space-charge impedance calculations, we consider the beam 
charge distribution in the form of a thin axisymmetric ring 
of radius $a$, $f(\vec{r}\,)=\delta(r-a)/(2\pi a)$.
Then the EF expansion coefficients are 
\begin{equation} 
f_{nm} = \delta_{n0}\frac{J_0(\mu_{0m}a/b)}{\sqrt{N^E_{0m}}} 
    \left \{ \begin{array}{c} 
      1 \\ 0 \end{array} 
    \right \}   \, ,                            \label{fnmc}
\end{equation} 
where $\delta_{nm}$ is the Kronecker symbol. The sum 
(\ref{sig}) becomes
\begin{eqnarray}
 S(f;\kappa) & = & \frac{1}{\pi} \sum^\infty_{m=1} 
  \frac{J^2_0(\mu_{0m}a/b)}{J^2_1(\mu_{0m})
    \left( \mu^2_{0m}+\kappa^2 b^2 \right)} \ \nonumber
\\ &=& \frac{I_0(\kappa a)}{2 \pi} 
 \left[ K_0(\kappa a) - I_0(\kappa a) 
 \frac{K_0(\kappa b)}{I_0(\kappa b)} \right] , \label{dsigc}      
\end{eqnarray}
where formulas \cite{BEII} were applied.

\section{Rectangular Chamber}

For a rectangular chamber of width $a$ and height $b$ the 
eigenvalues are $k_{nm} = \pi \sqrt{n^2/a^2+m^2/b^2}$ with 
$n,m = 1,2, \ldots$, and the normalized EFs are 
\begin{equation} 
e_{nm}(x,y) = \frac{2}{\sqrt{ab}}  \sin {\frac{\pi n x}{a}} 
 \sin {\frac{\pi m y}{b}}  \ ,                  \label{ernm}
\end{equation} 
with $0 \le x \le a$ and $0 \le y \le b$. Let a hole be located 
in the side wall at $x=a, \ y=y_h$. Assuming the beam displacement
$\vec{s}$ from the chamber axis $(a/2,b/2)$, we can perform one 
summation in Eq.~(\ref{enp}) to reduce the double sum into a 
fast-converging series
\begin{eqnarray}
e_\nu(\vec{s}\,;\kappa) & = & \frac{2}{b}
 \sum^\infty_{m=1} \sin [\pi m (\frac{1}{2}+\frac{s_y}{b})] 
 \sin (\frac{\pi m y_h}{b}) \times \nonumber 
\\ & & \quad \times \frac {\sinh [\pi u_m (1/2+s_x/a)]}
 { \sinh (\pi u_m) }        \,  ,                 \label{enpr}
\end{eqnarray}
where $u_m=\sqrt{m^2a^2/b^2+\kappa^2a^2/\pi^2}$.
For an on-axis beam, $s \to 0$, it can be simplified to
\begin{equation} 
 e_\nu(0;\kappa) = \frac{1}{b} \sum_{p=0}^{\infty} 
 \frac { (-1)^p \sin [\pi (2p+1)y_h/b ]}
 {\cosh [\pi u_{2p+1}/2] }    \, .               \label{enpr0}
\end{equation} 
In the relativistic limit, $\beta \to 1$, 
$\gamma \to \infty$, Eq.~(\ref{enpr0}) becomes
\begin{equation} 
 e_\nu(0;\kappa)\to e_\nu(0;0)\equiv \tilde{e}_{\nu}= \frac{1}{b} 
 \Sigma \left(\frac{a}{b},\frac{y_h}{b}\right) \,, \label{enrect}
\end{equation} 
where 
\begin{equation} 
 \Sigma (u,v) = \sum_{p=0}^{\infty} \frac { (-1)^p \sin 
 [\pi (2p+1)v]}{\cosh [\pi (2p+1) u/2] } \,,      \label{Sigma}
\end{equation} 
which coincides with the result \cite{KGS} for the rectangular
chamber. For some particular values of $v$, e.g.\ $v=1/2$, the 
sum $\Sigma (u,v)$ can be expressed in terms of the complete 
elliptic integrals; in general, it is easy to calculate the 
series numerically because of its fast (exponential) convergence. 
Its behavior versus $v$ for different values of the aspect 
ratio $u$ was plotted in Ref.~\cite{SK92}. 

The gradient of the field (\ref{enpr}) at the origin is required 
for calculating the transverse impedance. In the Cartesian 
co-ordinates $\{x,y\}$, it is
\begin{eqnarray}
\vec{d}\,(\kappa) & \equiv & \{d_x,d_y\} =
      \vec{\nabla} e_\nu(0;\kappa) =   \nonumber \\
& = & \frac{\pi}{a b} \left \{ \sum^\infty_{p=0} (-1)^p  
 \frac{u_{2p+1} \sin [\pi (2p+1)y_h/b]} 
  {\sinh (\pi u_{2p+1}/2) } ; \right.  \nonumber \\
 & & \qquad \left. \frac{a}{b} \sum^\infty_{p=0} (-1)^p  
 \frac{2p \, \sin (2\pi p y_h/b)} 
  {\cosh (\pi u_{2p}/2) } \right\} \,.        \label{grenpr} 
\end{eqnarray}
In the ultrarelativistic limit $u_m \to m a/b$, and gradient 
(\ref{grenpr}) becomes frequency-independent:
\begin{eqnarray}
\vec{d}\,(0) & = & 
 \frac{\pi}{b^2} \left \{ \sum^\infty_{p=0} (-1)^p  
 \frac{(2p+1) \sin [\pi (2p+1)y_h/b]} 
  { \sinh [\pi (p+1/2) a/b] } ; \right.  \nonumber \\
 & & \qquad \left. \sum^\infty_{p=0} (-1)^p  
 \frac{2p \, \sin (2\pi p y_h/b)} 
  {\cosh (\pi p a/b) } \right\} \,.             \label{grenpr1} 
\end{eqnarray}

For calculating the space-charge impedance of a centered 
rectangular hollow beam $\zeta a \times \zeta b$ in the 
beam pipe with rectangular cross section $a \times b$,
the EF-expansion coefficients are 
\begin{eqnarray}
f_{nm} & = & \frac{4  \sin (\pi n/2) \sin (\pi m/2) }
 {\pi^2 \zeta (a+b) \sqrt{a b}} \times         \label{fnmr}
\\ & & \times \! \left( \frac{a}{n} 
  \sin{\frac{\pi n \zeta}{2}} \cos{\frac{\pi m \zeta}{2}} 
  +  \frac{b}{m} \cos{\frac{\pi n \zeta}{2}}
  \cos{\frac{\pi m \zeta}{2}} \right ) .   \nonumber   
\end{eqnarray}
Substituting $f_{nm}$ into Eq.~(\ref{sig}) and performing one 
summation to reduce the resulting double sum to a series leads 
to Eq.~(\ref{sigr}).

\end{document}